

\def\pmb#1{\setbox0=\hbox{$#1$}%
\kern-.025em\copy0\kern-\wd0
\kern.05em\copy0\kern-\wd0
\kern-.025em\raise.0433em\box0 }
\def\ll{\left\langle}
\def\rr{\right\rangle}
\def\underarrow#1{\mathrel{\mathop{\longrightarrow}\limits_{#1}}}
\magnification=1200
\hoffset=-.1in
\voffset=-.2in

\vsize=7.5in
\hsize=5.6in
\tolerance 10000

\baselineskip 12pt plus 1pt minus 1pt
\pageno=0
\centerline{\bf ``PEIERLES SUBSTITUTION'' AND}
\smallskip
\centerline{{\bf CHERN--SIMONS QUANTUM MECHANICS}\footnote{*}{This
work is supported in part by funds
provided by the U. S. Department of Energy (D.O.E.) under contract
\#DE-AC02-76ER03069.}}
\vskip 24pt
\centerline{G. Dunne}
\vskip 12pt
\centerline{\it Department of Mathematics}
\centerline{and}
\centerline{\it Center for Theoretical Physics}
\centerline{\it Massachusetts Institute of Technology}
\centerline{\it Cambridge, Massachusetts\ \ 02139\ \ \ U.S.A.}
\vskip 12pt
\centerline{and}
\vskip 12pt
\centerline{R. Jackiw}
\vskip 12pt
\centerline{\it Center for Theoretical Physics}
\centerline{\it Laboratory for Nuclear Science}
\centerline{\it and Department of Physics}
\centerline{\it Massachusetts Institute of Technology}
\centerline{\it Cambridge, Massachusetts\ \ 02139\ \ \ U.S.A.}
\vskip 1.5in
\centerline{NOT FOR PUBLICATION}
\centerline{(in view of Girvin/Jach paper)}
\vfill
\centerline{ Typeset in $\TeX$ by Roger L. Gilson}
\vskip -12pt
\noindent GDandRJ\hfill April 1992
\eject
\baselineskip 24pt plus 2pt minus 2pt
\centerline{\bf ABSTRACT}
\medskip
An elementary derivation is given for the ``Peierles substitution'' used in
projecting dynamics in a strong magnetic field onto the lowest Landau level.
The projection of wavefunctions and the
ordering prescription for the projected Hamiltonian is explained.
\vfill
\eject
The ``Peierles substitution''$^1$ advances calculations for the following
problem.  Consider a charged particle with mass $m$ moving on the plane in a
constant magnetic field $B$ perpendicular to the plane, and also subject to
planar forces arising from the potential $V(x,y)$.  Motion along the $z$-axis
(along $B$) is ignored and it is assumed that $B$ is sufficiently strong and
$V$ sufficiently weak so that the lowest Landau level retains its identity in
the presence of $V$.   The question one wants to answer is how the
energy-degenerate states in the lowest Landau level are modified by the
interaction with $V$.  The result in the approximation of strong $B$ and weak
$V$ is that the energy eigenvalues become
$$E = {B\over 2m} + \epsilon_n \eqno(1)$$
where $B/2m$ is the lowest Landau level energy [we set $\hbar$, $c$ and
particle charge to unity] while $\epsilon_n$ are eigenvalues of the operator
obtained from $V(x,y)$ by the Peierles substitution,
$$V\left(p, q\right) |n\rangle = \epsilon_n|n\rangle \eqno(2)$$
where $p$ and $q$ are canonically conjugate (with $1/B$ playing the role of
Planck's constant).
$$i[p,q]={1\over B}\eqno(3)$$

The Peierles substitution has been used for over half a century; a recent
application is to the Azbel'$^2$--Hofstadter$^3$ problem where $V(x,y) = V_0
 \cos x + V_0 \cos y$.  But its justification$^4$ remains a ``most
difficult step.''$^3$

In this Letter we draw on our experience with Chern--Simons quantum
mechanics$^5$ to give an immediate derivation of the Peierles substitution,
and to answer two immediate questions: first, how do the wavefunctions
$\psi(x,y)$ which depend on two commuting coordinates $x$ and $y$ reduce to
wavefunctions depending only on the single coordiante $q$?  Second, how does
one resolve the operator ordering ambiguity inherent in (2) when the potential
$V(x,y)$ becomes a function of the non-commuting operators $p$ and $q$?  We
answer these questions and explicitly verify our prescription when $V$ is
rotationally symmetric: $V=V(x^2 + y^2)$.

Our derivation relies on the following fact about quantizing a system
governed by
a Lagrangian that is {\it linear\/} in time derivatives.  Let $\xi^i$
($i=1,2$) be a two-component quantity with Lagrangian
$$L = {g\over 2} \sum_{ij}\dot\xi^i \epsilon_{ij} \xi^j - V(\xi) \eqno(4)$$
Here $g$ is a constant, $\epsilon_{ij}$ the anti-symmetric tensor
$\epsilon_{ij} = \left(\matrix{ \phantom{-}0&1\cr-1&0\cr}\right)_{ij}$, and $V$
a function depending on $\xi^i$ but not on its time-derivative $\dot\xi^i$.
Although lacking the
usual kinetic term, quadratic in time derivatives, (4) gives rise to
non-singular
equations of motion, which may be quantized in a Hamiltonian framework,
provided the fundamental quantum commutator is taken as
$$\left[ \xi^i, \xi^j\right] = {i\over g} \epsilon^{ij} \ \ .\eqno(5)$$
While (5) is a standard result$^6$ about systems like (4), it is not
well-known, hence in the Appendix we sketch a derivation.

Consider now the magnetic problem described by the Lagrangian
$$L_m = {1\over 2} m\dot{\bf r}^2 + {B\over 2}{\bf r}\times \dot{\bf r} -
V({\bf
r}) \eqno(6)$$
Dynamics is confined to the plane, ${\bf r} = (x,y)$, the $z$-coordinate is
suppressed and the vector potential ${\bf A}$ giving rise to the
constant magnetic field $B = \pmb{\nabla}\times{\bf A}$ is taken in the
rotationally symmetric gauge
$A^i(x) = - \epsilon^{ij} r^j {B\over 2}$.
The Hamiltonian corresponding to (6) is the familiar expression
$$H_m ({\bf p}, {\bf r}) = {1\over 2m} \left( {\bf p} - {\bf A}({\bf
r})\right)^2 + V({\bf r}) \eqno(7)$$
In the absence of $V$, the spectrum is well-known: the infinitely degenerate
lowest Landau level has energy $B/2m$ and the higher levels are separated by
gaps of magnitude proportional to $B/m$.  When $V$ enters weakly and $B$
is strong, the degeneracy is lifted, but the pattern remains unchanged,
and {\it the lowest level may be isolated by setting $m$ to zero\/}, provided
the energy eigenvalue $B/2m$ is subtracted (``renormalized'').  {\it Thus the
structure of the lowest level is governed by the $m=0$ limit\/}.

[The analogy to gauge theories with Chern--Simons terms is now evident: the
kinetic and potential terms in (6) are point-particle analogues of the
conventional Maxwell/Yang--Mills Lagrangian; the second, magnetic term models
the field theoretic Chern--Simons interaction.  Setting $m$ to zero in the
quantum mechanics problem parallels the passage from a topologically massive
gauge field theory (with the Maxwell/Yang--Mills {\it and\/}
 Chern--Simons terms) to a
pure Chern--Simons field theory.  Indeed Ref.~[5] is devoted to an
illustration of these
field-theoretic issues in a quantum mechanical context.  But here now we
pursue the actual physical relevance of the quantum mechanical model.]

The $m\to0$ limit may be discussed in the Lagrangian (6) or Hamiltonian (7)
framework.  In view of the previously stated result (4) and (5), the
Lagrangian discussion directly leads to the conclusion that the symplectic
structure of the reduced Lagrangian
$$L_0 = {B\over 2} {\bf r}\times \dot{\bf r} - V({\bf r}) \eqno(8)$$
enforces the commutation relation
$$i\left[ r^i, r^j\right] = {1\over B}\epsilon^{ij} \eqno(9)$$
and the reduced Hamiltonian $H_0$, which is obtained from $L_0$ by the usual
Legendre transform
$$H_0 = {\partial L_0\over \partial\dot{\bf r}} \cdot \dot{\bf r} - L_0 =
V({\bf r})\ \ , \eqno(10)$$
becomes evaluated with non-commuting arguments in view of (9).  An
alternative, more complicated, derivation starts with $H_m$ of (7):
the limit $m=0$ is achieved by imposing quantum mechanically the constraint
$$0 = {\bf p} - {\bf A} \equiv m\dot{\bf r}\eqno(11)$$
$H$ reduces to $H_0$ and (9) emerges as a consequence of constrained
quantization.$^{5,\,6}$

We have thus arrived at a derivation of the Peierles substitution: the
perturbed
structure of the lowest Landau level is governed by the effective Hamiltonian
$$H_0 = V\left(p,q\right) \eqno(12)$$
where the non-commutativity of the arguments follows from (9) (expressed as in
(3)).

When $V=V(r^2) = V(p^2+q^2)$ it is temptng to write the eigenvalues of $H_0$
as
$$\epsilon_n \sim V\left( {2\over B}\left( n + {1\over 2}\right)\right)
\eqno(13)$$
where ${2\over B}\left(n+{1\over 2}\right)$
are the harmonic oscillator eigenvalues.  However, this
naive substitution ignores operator ordering issues, which produce
``corrections'' to (13).

To study the operator ordering
 we first describe the behavior of wavefunctions when
the limit $m\to 0$ is taken. Note that in the limit, phase space is reduced
from four dimensions
$(p_x, p_y, x,y)$ to two $(p,q)$; this is seen also as a consequence of the
constraint (3).   Wave functions depend on half the
phase-space variables: two before the reduction and one after, but if
normalization is maintained, one
argument of $\psi_m(x,y)$ cannot simply {\it disappear\/}
 when $m$ vanishes.  Detailed
analysis$^5$ gives the following story.  In quantizing the reduced theory, the
holomorphic polarization is chosen for wavefunctions:$^7$ the single
combination of the two phase space variables, on which wavefunctions are
taken to depend, is the non-Hermitian combination $\sqrt{B/2}\,(x+iy)$.  Thus
we form the operators
$$a \equiv \sqrt{{B\over 2}} (x-iy)\ \ ,\qquad a^\dagger \equiv \sqrt{{B\over
2}} (x+iy)\ \ ,\qquad [a,a^\dagger] = 1 \eqno(14)$$
Coherent states provide a basis
$$\langle \alpha |a^\dagger = \langle \alpha|\alpha \eqno(15)$$
and states $|\psi\rangle$ are described by
wavefunctions that depend on $\alpha$.
$$\ll \alpha|\psi\rr = \psi(\alpha) \eqno(16)$$
The operator $a^\dagger$ acts on these functions by multiplication, a by
differentiation; the adjoint relationship between the two is maintained by
virtue of a non-trivial measure.
$$\eqalign{\ll \alpha |a^\dagger|\psi\rr &= \alpha \psi(\alpha) \cr
\ll \alpha |a|\psi\rr &= {d\over d\alpha} \psi(\alpha) \cr}\eqno(17)$$
$$\eqalign{&{1\over 2\pi i} \int d\alpha^*d\alpha e^{-|\alpha|^2}
|\alpha\rangle \langle \alpha| = I \cr
&{1\over 2\pi i} d\alpha^*d\alpha \equiv {B\over 2\pi} dx\,dy \ \ .\cr}
\eqno(18)$$
One is also interested in number states
$$a^\dagger a |n\rangle = n|n\rangle \eqno(19\hbox{a})$$
which are described within the holomorphic representation by
$$\ll \alpha|n\rr = {\alpha^n\over \sqrt{n!}}\ \ .\eqno(19\hbox{b})$$

The phase-space reductive $m\to 0$ limit onto the lowest Landau level, when
applied to wavefunctions, works in
the following manner.  The $(x,y)$-dependence in the finite-$m$
eigenfunctions of $H_m$ for the lowest Landau level, are presented in terms of
complex variables $z = \sqrt{B/2} \, (x+iy)$, $z^*=\sqrt{B/2}\,(x-iy)$
$$\psi_m = \psi_m (z,z^*)\eqno(20\hbox{a})$$
When $m$ is set to zero, the limiting form becomes
$$\lim\limits_{m\to 0} \psi_m (z,z^*) = \left({B\over 2\pi}\right)^{1/2}
e^{-{1\over 2}|z|^2} \psi(z)\eqno(20\hbox{b}) $$
with $\psi$ being a holomorphic eigenfunction of $H_0$. Note that the
probability densities transform properly so that the norm is maintained.
$$d^2r \left| \psi_m({\bf r})\right|^2 \underarrow{m\to 0} {B\over 2\pi}
dx\,dy\, e^{-|z|^2} |\psi(z)|^2 = {d\alpha\, d\alpha^*\over 2\pi i}
e^{-|\alpha|^2} |\psi(\alpha)|^2 \eqno(21)$$
(All this
may be easily and explicitly checked for the solvable case when $V$ is a
harmonic oscillator potential.$^5$)

The ordering prescription for $H_0=V$ must be such that the matrix elements
$\epsilon=\ll n|V|n\rr$ of $V$, when computed before the phase-space reduction
as integrals over the {\it function\/} $V(x,y)$, coincide with the evaluation
{\it after\/} reduction, when $V$ becomes an {\it operator\/} between the
number states $|n\rangle$ in (19a).  This prescribes that the operator $H_0$
be expressed as the {\it anti-normal ordered\/} ({\it i.e.\/} all $a$'s to the
left, $a^\dagger$'s to the right) form of $V$ when $V(x,y)$ is written as
$V(z,z^*)$ and $z^*$ is replaced by $a$ and $z$ by $a^\dagger$.  (To see the
need for anti-normal ordering, note that the coherent state resolution of the
identity in (18) implies that
$a^k \left(a^\dagger\right)^\ell = \int {dz\,dz^*\over 2\pi i} e^{-|z|^2}
|z\rangle \left( z^*\right)^k z^\ell \langle z|$.)

To verify this explicitly, we consider a potential $V = V(r^2)$ as a function
of $r^2 = p^2 + q^2 \sim {2\over B} a^\dagger a$, a potential where the tilde
indicates that the ordering is yet to be performed.  Then $mH_m$ may be
written as
$$mH_m ={1\over 2} ({\bf p}-{\bf A})^2 + mV\eqno(22)$$
and the energies $\epsilon_n$ of (1) can be evaluated by lowest order
perturbation theory in $m$.
The ground state of the unperturbed Hamiltonian
 $({\bf p} - {\bf A})^2/2$ carries energy $B/2$; it is infinitely degenerate
with wavefunctions
$$u_n ({\bf r}) = {1\over \sqrt{\pi}} \left( {B\over 2}\right)^{(1+n)/2}
{z^n\over\sqrt{n!}} e^{-{B\over 4}|z|^2}\ \ ,\qquad n = 0,1,\ldots  \eqno(23)$$
Degenerate perturbation theory must be employed, but fortunately a
rotationally symmetric $V$ is already diagonal in the representation (23).
Thus we find
$$\epsilon_n = \left( {B\over 2}\right)^{1+n} \int^\infty_0 dr^2\, r^{2n} \,
V(r^2) e^{-{B\over 2} r^2} \eqno(24)$$
If $V$ is represented as
$$V(r^2) = \int d\sigma\, e^{-\sigma r^2} v(\sigma) \eqno(25)$$
it follows that
$$\epsilon_n =\int d\sigma \left( 1 + {2\over
B}\sigma\right)^{-1-n} v(\sigma)\eqno(26)$$
On the other hand, in the reduced problem we have the operator Hamiltonian
$$V\sim \int d\sigma\, e^{-{2\over B}\sigma a^\dagger a}v(\sigma)
\eqno(28\hbox{a})$$
To anti-normal order, we expand the exponential
$$\eqalign{V &= \int d\sigma\, v(\sigma) \sum^\infty_{N+0} {1\over N!} \left( -
{2\sigma\over B}\right)^N (a)^N (a^\dagger)^N\cr
&= \int d\sigma\, v(\sigma) \left(1+ \sum^\infty_{N=1} {1\over N!} \left(
- {2\sigma\over B}\right)^N
\left( a^\dagger a+N\right) \left( a^\dagger a + N-1\right) \ldots
\left( a^\dagger a+2\right) \left( a^\dagger a+1\right)\right)
 \cr}\eqno(28\hbox{b})$$
The number states $|n\rangle$ are eigenstates with eigenvalue
$$\epsilon_n = \int d\sigma\, v(\sigma) \sum^\infty_{N=0} {1\over N!} \left( -
{2\sigma\over B}\right)^N {(N+n)!\over n!} \eqno(29)$$
Summing the series reproduces (26).  These expressions, (26) and (29),
encapsulate all the ``corrections'' to the naive replacement in (13).
\bigskip
\centerline{\bf ACKNOWLEDGEMENT}
\medskip
Dimensional reduction of phase space in the presence of strong magnetic fields
has recently also been studied by Iso, Karabali, Sakita, Sheng and Su,$^8$
as well as by Levit and Siran.$^9$  We thank some of these
authors for discussion.
Levit and Sivan give a semi-classical (WKB) analysis of ordering corrections
to $H_0$; their prescription in general disagrees with ours, agreement exists
only for the first two terms of a large-$n$ expansion.  We are grateful to
C.~Callan and D. Freed for acquainting us with the Peierles substitution and M.
Berry for suggesting the perturbative calculation.
\goodbreak
\bigskip
\centerline{\bf NOTE ADDED}
\medskip
After completing this work, we searched the literature and discovered that
most of our points were made almost a decade ago by S.~Girvin and T.~Jach,
{\it Phys. Rev. D\/} {\bf 29}, 5617 (1984); see also S.~Kivelson, C.~Kallin,
D.~Arovas and J.~Schrieffer, {\it Phys. Rev. B\/} {\bf 36}, 1620 (1987).
\vfill
\eject
\centerline{\bf REFERENCES }
\medskip
\item{1.}R. Peierles, {\it Z. Phys.\/} {\bf 80}, 763 (1933).
\medskip
\item{2.}M. Azbel', {\it Zh. Eksp. Teor. Fiz.\/} {\bf 46}, 939 (1964).
[English translation: {\it Sov. Phys. JETP\/} {\bf 19}, 634 (1964).]
\medskip
\item{3.}D. Hofstadter, {\it Phys. Rev. B\/} {\bf 14}, 2239 (1976).
\medskip
\item{4.}J. Luttinger, {\it Phys. Rev.\/} {\bf 84}, 814 (1951); W. Kohn, {\it
Phys. Rev.\/} {\bf 115}, 1460 (1959); G. Wannier, {\it Rev. Mod. Phys.\/} {\bf
34}, 645 (1962); E. Blount, {\it Phys. Rev.\/} {\bf 126}, 1636 (1962).
\medskip
\item{5.}G. Dunne, R. Jackiw and C. Trugenberger, {\it Phys. Rev. D\/} {\bf
41}, 661 (1990).
\medskip
\item{6.}L. Faddeev and R. Jackiw, {\it Phys. Rev. Lett.\/} {\bf 60}, 1692
(1988).
\item{7.} See {\it e.g.\/} L. Faddeev, in {\it Methods in Field Theory \/}, R.
Balian and J. Zinn--Justin, eds. (North-Holland, Amsterdam/World Scientific,
Singapore, 1985).
\medskip
\item{8.} S. Iso, University of Tokyo preprint UT-579 (February 1991);
B. Sakita, D. N. Sheng and Z.-B. Su, {\it Phys. Rev. B\/} {\bf 44},
11510 (1991); D. Karabali and B. Sakita, {\it Intl. Jnl. Mod. Phys. A\/} {\bf
6}, 5079 (1991); S. Iso, D. Karabali and B. Sakita, CCNY preprint HEP-92/1
(January, 1992).
\medskip
\item{9.}N. Siran and S. Levit, Weizmann preprints WIS-91/31, 92/1, 92/6.
\vfill
\eject
\centerline{\bf APPENDIX}
We consider the following Lagrangian that governs dynamics for a multiplet
of variables $\xi^i$ and is linear in the velocities $\dot\xi^i$.
$$L = a_i(\xi) \dot\xi^i - V(\xi) \eqno(\hbox{A.1})$$
Here $a_i$ and $V$ are functions of $\xi^i$, but not of $\dot\xi^i$, and the
summation convention is used.  $L$ of (A.1) is a generalization of
(4), where $a_i (\xi) = {g\over 2}\epsilon_{ij} \xi^j$.  Although (A.1)
lacks the usual  kinetic term, quadratic in velocities, it is in fact the
generic form for a Lagrangian.  The point is that familiar Lagrangians like
$$L = {1\over 2} m \dot q^2 - V(q) \eqno(\hbox{A.2a})$$
may be expressed as
$$L(p,q) = p\dot q - H (p,q) \ \ ,\qquad H(p,q) = {p^2\over 2m} + V(q)
\eqno(\hbox{A.2b})$$
and the Euler-Lagrange equations for $L(p,q)$, with $p$ and $q$ taken as
Lagrangian variables, reproduce the equations of motion of (A.2a).  But (A.2b),
is of the form (A.1) with $\xi^i = {p\choose q}$.

The equations of motion that follow from (A.1) read
$$f_{ij}(\xi)\dot\xi^j = {\partial V(\xi)\over
\partial\xi^i}\eqno(\hbox{A.3})$$
$$f_{ij} (\xi) \equiv {\partial\over\partial \xi^i} a_j (\xi) - {\partial
\over \partial\xi^i} a_i(\xi) \eqno(\hbox{A.4})$$
Next, we assume that the matrix $f_{ij}$ possesses an inverse $f^{ij}$
--- if it does
not, the subsequent development is more complicated,$^6$ but unnecessary in
our case (4).  Then (A.3) implies
$$\dot\xi^i = f^{ij}{(\xi)} {\partial V(\xi)\over \partial \xi^j}
\eqno(\hbox{A.5})$$
The Hamiltonian for (A.1) is constructed by the usual Legendre
transform
$$H = {\partial L\over \partial \dot\xi^i} \dot\xi^i - L = V(\xi)
\eqno(\hbox{A.6})$$
Hence if the aim is to reproduce (A.5) by bracketing $\xi^i$ with the
Hamiltonian, it must be that the $[\xi^i,\xi^j]$ bracket isnon-vanishing since
$H=V(\xi)$ depends only on $\xi^i$.  Evidently,
$$ i \left[ V(\xi), \xi^i\right] = {\partial V(\xi_i)\over \partial
\xi^j} i \left[ \xi^j, \xi^i\right] =\dot\xi^i\eqno(\hbox{A.7})$$
and (A.5) is regained provided
$$\left[ \xi^i, \xi^j\right] = i\, f^{ij} (\xi) \eqno(\hbox{A.8})$$
This gives the derivation of (5).

\par
\vfill
\end